\documentclass[aip,reprint, jmp, amsmath, amssymb]{revtex4-1}
\usepackage{graphicx}
\usepackage{dcolumn}
\usepackage{amsmath}
\usepackage{bm}

\begin{document}

\title{Optimal preselection and postselection in weak measurements for observing photonic spin Hall effect}
\author{Xinxing Zhou}
\author{Xing Li}
\author{Hailu Luo}
\email{hailuluo@hnu.edu.cn}
\author{Shuangchun Wen}
\email{scwen@hnu.edu.cn} \affiliation{Key Laboratory for
Micro-/Nano-Optoelectronic Devices of Ministry of Education, College
of Physics and Microelectronic Science, Hunan University, Changsha
410082, People's Republic of China}
\date{\today}

\begin{abstract}
Photonic spin Hall effect (SHE) holds great potential applications
in precision metrology. How to obtain a high measurement precision
is an important issue to detect the photonic SHE. In this letter, we
propose using optimal preselection and postselection in weak
measurements to enhance the measurement precision. We find that the
maximum weak value and pointer shift can be obtained with an optimal
overlap of preselection and postselection states. These findings
offer the possibility for improving the precision of weak
measurements and thereby have possible applications for accurately
characterizing the parameters of nanostructures.
\end{abstract}

\maketitle

After first proposed by Aharonov, Albert, and Vaidman
(AAV)~\cite{Aharonov1988} in 1988, the weak measurements based on
preselection and postselection states has been a promising method
for helping us in investigating fundamental questions of quantum
mechanics~\cite{Ritchie1991,Pryde2005,Dressel2012}. The idea of weak
measurements can be described as follows: if we initially select the
quantum system with a well-defined preselection state, the
corresponding large expectation values can be obtained with a
suitable postselection state, which makes the eigenvalues to be
clearly distinguished. Recently, the weak measurements has been
useful for precision measurement such as detecting very small
transverse beam deflections~\cite{Dixon2009}, observing photonic
spin Hall effect
(SHE)~\cite{Hosten2008,Qin2009,Gorodetski2012,Luo2011}, determining
the average trajectories of single photons~\cite{Kocsis2011}, direct
measurement of the quantum wavefunction~\cite{Lundeen2011}, and
measuring ultrasmall time delays of light~\cite{Strubi2013}.

The photonic SHE is attributed to spin-orbit coupling and manifests
itself as spin-dependent
splitting~\cite{Onoda2004,Bliokh2006,Aiello2008,Haefner2009,Luo2009,Hermosa2011,Shitrit2011,Yin2013}.
The photonic SHE is sensitive to the variations of physical system's
states and holds great potential applications in precision
metrology, such as probing spatial distributions of electron spin
states~\cite{Menard2009}, measuring the thickness of nanometal
film~\cite{Zhou2012a}, identifying graphene layers~\cite{Zhou2012b},
and detecting the axion coupling in topological
insulators~\cite{Zhou2013}. The spin-dependent splitting of photonic
SHE in these systems is just a few tens of nanometers and the weak
measurement method is usually used to probe this phenomenon.
However, in the process of weak measurements for probing the
photonic SHE, how to get the large amplified factor is an important
issue for enhancing the measurement precision.

In this letter, we propose using optimal preselection and
postselection in weak measurements to detect the photonic spin Hall
effect (SHE) for obtaining the maximum outcome. We consider the
regime of preselection and postselection being almost orthogonal. It
is found that, for a fixed incident angle, the maximum amplified
factor and pointer shift (amplified displacement of photonic SHE)
can be obtained with a corresponding optimal overlap of preselection
and postselection states. Therefore, we can significantly improve
the precision of weak measurements for probing the photonic SHE. We
also find that, under the orthogonal condition of preselection and
postselection states, the amplified factor and amplified shift can
not be arbitrarily large and, on the contrary, they turn out to be
zero. The experimental results agree well with our theoretical
discussions.

In general, the amplified factor is corresponding to the so-called
weak value which establishes the relationship between the observable
and the shifts in measurement pointer's mean position and mean
momentum~\cite{Jozsa2007}
\begin{equation}
A_{w}=\frac{\langle
\psi_{2}|\mathbf{\hat{A}}|\psi_{1}\rangle}{\langle
\psi_{2}|\psi_{1}\rangle}.\label{weak value}
\end{equation}
$\psi_{1}$ and $\psi_{2}$ denote the preselection and postselection
states. When $\langle \psi_{2}|\psi_{1}\rangle$$\rightarrow$0, the
weak value might become arbitrarily large. However, arbitrarily
large results in photonic SHE can not be obtained and we need to
choose an appropriate preselection and postselection states for
obtaining the maximum weak value. Some theoretical works have been
done to discuss this phenomena~\cite{Geszti2010,Wu2011,Zhu2011},
however, there lacks specific experiment and the photonic SHE is not
considered.

\begin{figure}
\includegraphics[width=8cm]{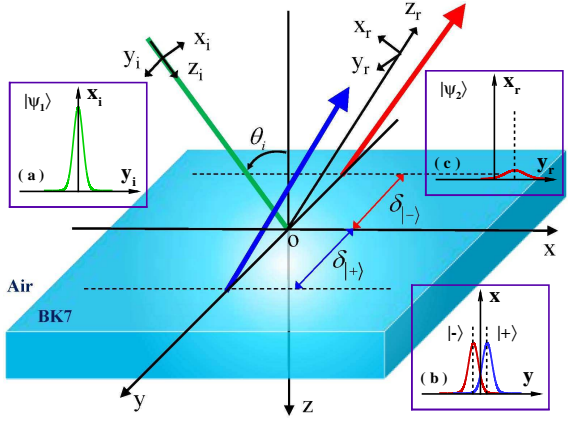}
\caption{\label{Fig1} (Color online) Schematic of photonic SHE on a
BK7 prism interface and the corresponding weak measurement process.
A linearly polarized beam reflects on the prism and then splits into
left- and right-handed circularly polarized light, respectively.
$\delta_{|+\rangle}$ and $\delta_{|-\rangle}$ denote the transverse
shifts of left- and right-handed circularly polarized components.
Here, $\theta_{i}$ is the incident angle and the insets show the
three steps of weak measurements: (a) preselection, (b) weak
coupling between observable and meter, and (c) postselection. }
\end{figure}

Figure~\ref{Fig1} schematically draws the photonic SHE of light beam
reflection from a planar interface and the corresponding weak
measurement process. The incident polarization states are chosen as
$|H\rangle$ and $|V\rangle$. This polarization selection can be seen
as the preselection process in weak measurements, which is discussed
in the following. In the spin basis, the horizontal and vertical
polarization states can be expressed as
$|H\rangle=(|+\rangle+|-\rangle)/{\sqrt{2}}$ and
$|V\rangle=i(|-\rangle-|+\rangle)/{\sqrt{2}}$. In the spin basis,
the states of reflected beam can be obtained:
\begin{equation}
|H\rangle\rightarrow\frac{r_{p}}{\sqrt{2}}\left[\exp(+ik_{ry}\delta^{H}_{r})|+\rangle+\exp(-ik_{ry}\delta^{H}_{r})|-\rangle\right]\label{H
spectrum},
\end{equation}
\begin{equation}
|V\rangle\rightarrow\frac{ir_{s}}{\sqrt{2}}\left[-\exp(+ik_{ry}\delta^{V}_{r})|+\rangle+\exp(-ik_{ry}\delta^{V}_{r})|-\rangle\right]\label{V
spectrum}.
\end{equation}
In the above equations,
$\delta^{H}_{r}=(1+r_{s}/r_{p})\cot\theta_{i}/k_{0}$,
$\delta^{V}_{r}=(1-r_{p}/r_{s})\cot\theta_{i}/k_{0}$.

The photonic SHE manifests for the spin-dependent splitting of left-
and right-handed circularly polarized components. Here we only
consider the spin separation in the y direction (transverse shift).
In the following, we calculate the shifts of these two spin
components. The wavefunction of reflected photons is composed of the
packet spatial extent $\phi(k_{ry})$ and the polarization
description $|H,V\rangle$:
\begin{equation}
|\Phi^{H,V}\rangle=\int dk_{ry}
\phi(k_{ry})|k_{ry}\rangle|H,V\rangle\label{inital}.
\end{equation}
After photons reflection from the interface, the initial state
$|\Phi_{inital}^{H,V}\rangle$ evolve into the final state
$|\Phi_{final}^{H,V}\rangle$. As a result of spin-orbit coupling,
the displacements of the two spin components compared to the
geometrical-optics prediction are given by
\begin{equation}
\delta_{|\pm\rangle}^{H,V}=\frac{\langle
\Phi^{H,V}|i\partial_{\mathbf{k_\perp}}|{\Phi^{H,V}}\rangle}{\langle
\Phi^{H,V}|\Phi^{H,V}\rangle}.\label{BCII}
\end{equation}
Here, we suppose the $\phi(k_{ry})$ is a Gaussian wave function.

In the process of quantum weak measurements, the property observable
of a system is first coupled to the meter (measuring device), and
then the information about the state of the observable is read out
from the meter. In the case at hand, the detection of photonic SHE
induced transverse shifts is actually equivalent to a quantum
measurement of the spin degree of freedom along the central
propagation direction corresponding to the observable
$\hat{\sigma}_{3}$, with the transverse spatial distribution serving
as the meter~\cite{Hosten2008}. This can be done through three
steps. Firstly, the system is prepared with a fixed preselection
state $\psi_{1}$. In the present work, the incident polarization
states are chosen as $|H\rangle$ and $|V\rangle$. And then, with the
weak interaction, the observable (left- or right-handed circularly
polarized component) is coupled to the transverse spatial
distribution of the Gaussian wave function with the Hamiltonian
interaction. Finally, an enhanced shift in the meter distribution
can be obtained with the suitable postselection state
$|\psi_{2}\rangle=|V\pm\Delta\rangle$ or $|H\pm\Delta\rangle$ of the
observable. Here, the $\Delta\ll1$ is a small angle. The last step
can also be seen as the strong measurement. When the $\Delta$ is
close to zero, the preselection and postselection states are almost
orthogonal.

Our experimental setup is similar to that in Ref.~\cite{Luo2011} and
the detail equipment description and experimental analysis can be
found. Our sample is an usual BK7 prism. The amplified displacements
can be obtained through the preselection and postselection process,
and so we can get the relationship between the final position of the
meter and the observable $\mathbf{\hat{A}}$ by the weak value
$A_{w}$. A Gauss beam generated by He-Ne laser is firstly focused by
the lens (L1) and experiences preselection in the state
$|\psi_{1}\rangle$=$|H\rangle$ or $|V\rangle$ with the polarizer P1.
When the light beam reflects from the prism interface, the photonic
SHE happens allowing for the left- and right-handed circularly
polarized components splitting in the y direction. This process is
the weak interaction allowing for the coupling between the
observable and the meter. And then the beam passes through the
second polarizer P2 preparing for the postselection state
$|\psi_{2}\rangle=|V\pm\Delta\rangle$ or $|H\pm\Delta\rangle$. We
can obtain the reflected field at the plane of $z_{r}$. At the
surface of the second polarizer, the two circular polarization
components experience destructive interference making the enhanced
shift $\delta_{w}^{H,V}$ in the meter much larger than the initial
one $\delta_{|\pm\rangle}$. Calculating the reflected field
distribution yields the amplified shifts of photonic SHE. After
passing through the second lens (L2), a CCD is used to capture the
optical signal and measure the amplified shifts. The process
discussed above is called the weak value amplification and $\Delta$
is the postselection angle.

\begin{figure}
\includegraphics[width=8.5cm]{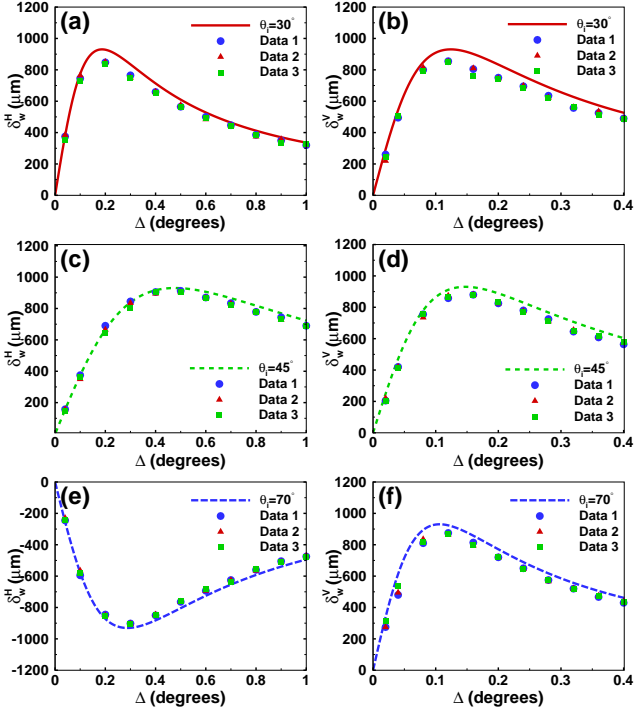}
\caption{\label{Fig2} (Color online) The amplified shifts of
photonic SHE changing with the overlap of the preselection and
postselection states in the case of horizontal (left column) and
vertical polarizations (right column). Here the incident angles are
chosen as $\theta_{i}$=$30^{\circ}$, $45^{\circ}$, and $70^{\circ}$.
The lines represent the theoretical results. The circles, squares,
and triangles show the experimental data. }
\end{figure}

In terms of weak value amplification, the preselection and
postselection states $|\psi_{1}\rangle$ and $|\psi_{2}\rangle$
determine the weak value $A_{w}$ of the photon helicity. We should
note that the imaginary weak value also corresponds to a shift of
the meter in momentum space, which leads to the possibility of even
larger enhancements following the beam free
evolution~\cite{Hosten2008}. This process can be seen as propagation
amplification that produces the amplified factor F. The propagation
amplified factor depends on the initial state of the beam and the
degree of its free evolution before the observer. Therefore, under
this condition, we can also get the final shift of the meter as the
modified weak value $A_{w}^{mod}=|A_w|F$. From the calculation, we
can obtain the modified weak value
\begin{eqnarray}
A_{w}^{Hmod} &=&
\frac{z_{r}k_{0}r_{p}^{2}\sin(2\Delta)}{(r_{p}+r_{s})^{2}\cot^{2}\Delta\cot^{2}\theta_{i}+2k_{0}z_{R}r_{p}^{2}\sin^{2}\Delta}
\label{H factor},
\end{eqnarray}
\begin{eqnarray}
A_{w}^{Vmod} &=& \frac{z_{r}k_{0}
r_{s}^{2}\sin(2\Delta)}{(r_{p}+r_{s})^{2}\cot^{2}\Delta\cot^{2}\theta_{i}+2k_{0}z_{R}r_{s}^{2}\sin^{2}\Delta}
\label{V factor}.
\end{eqnarray}

\begin{figure}
\includegraphics[width=8.5cm]{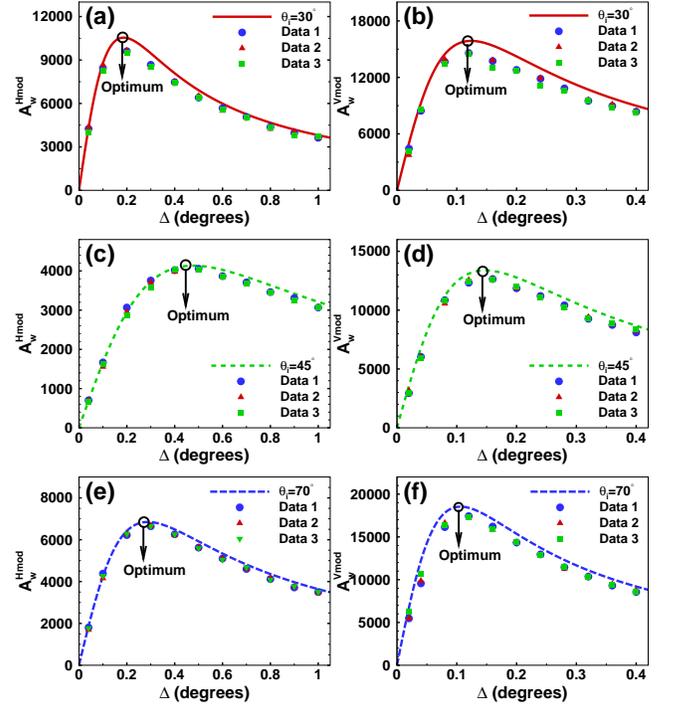}
\caption{\label{Fig3} (Color online) The theoretical and
experimental results for selecting the maximum weak value in weak
measurements under the condition of preselection and postselection
states being almost orthogonal. Here the incident angles are
selected as $\theta_{i}$=$30^{\circ}$, $45^{\circ}$, and
$70^{\circ}$. The incident states are horizontal (left column) and
vertical polarizations (right column). The circles, squares, and
triangles denote the experimental data corresponding to the
theoretical lines. }
\end{figure}

Figure~\ref{Fig2} shows the amplified displacements of photonic SHE
changing with the incident angle $\theta_{i}$ and the degree of
overlap of preselection and postselection states (described by the
postselection angle $\Delta$). Here, the incident angles are chosen
as three fixed values $\theta_{i}$=$30^{\circ}$, $45^{\circ}$, and
$70^{\circ}$. We repeat the experiment for three times. The shifts
are measured in the case of both $|H\rangle$ and $|V\rangle$
polarizations. Combining with the amplified shifts, we can obtain
the amplified factor (modified weak value) in the process of weak
measurements as shown in Fig.~\ref{Fig3}. We surprisingly find that
the weak value and amplified displacements can not increase
arbitrarily under the orthogonal condition of preselection and
postselection states and, on the contrary, it turns out to be zero.
Instead, there exists the maximum weak value and amplified
displacements with a corresponding optimal overlap of preselection
and postselection states. With the maximum weak value, we can get
the high measurement precision for probing the photonic SHE. We note
that the similar behavior of weak value is investigated in
solid-state qubits in which the conditioned average of a finite
strength measurement cross through zero at the orthogonality point,
and achieve both maximum and minimum values away from this
point~\cite{Williams2008}. In fact, to have the divergence of the
weak value, the ordering of the small parameters in weak measurement
process must be obeyed. If the ordering of the small parameters is
reversed, entirely new physical behavior is expected, and in fact
the inverse weak value can appear~\cite{Starling2010a,Starling2010}.

\begin{figure}
\includegraphics[width=6cm]{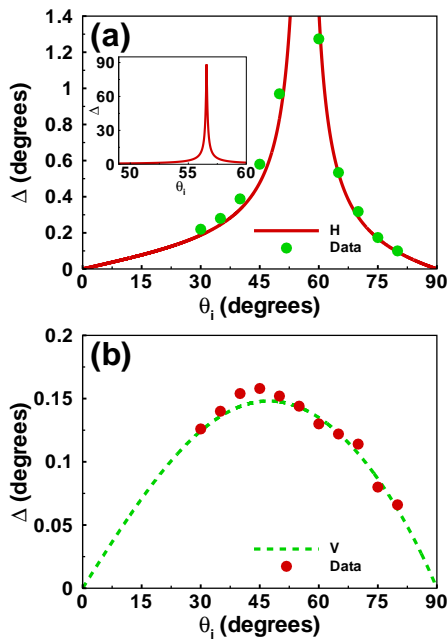}
\caption{\label{Fig4} (Color online) The relationship between
incident angle $\theta_{i}$ and postselection angle $\Delta$ for
obtaining the maximum amplified factor. The incident states are (a)
horizontal and (b) vertical polarizations. The circles show the
experimental data and the lines denote the theoretical value. The
inset describes the sharp value of postselection angle which is not
shown in the figure. }
\end{figure}

From Eqs.~(\ref{H factor}) and~(\ref{V factor}), using
$\frac{\partial A_{w}^{mod}}{\partial\Delta}$=0, we can get the
relationship between incident angle $\theta_{i}$ and postselection
angle $\Delta$ for obtaining the maximum amplified factor. The
results can be seen from Fig.~\ref{Fig4} and the preselection states
are chosen as $|H\rangle$ and $|V\rangle$ polarizations. We should
note that the incident angles can relate to the degree of weak
interaction in weak measurements. So, for a fixed incident angle,
there exists a maximum amplified factor corresponding to a optimal
overlap of preselection and postselection states. In the case of
$|H\rangle$ state, a sharp peak [see inset of Fig.~\ref{Fig4}(a)]
appears when the incident angle is close to Brewster angle. Here, we
have considered the tiny in-plane spread of wave-vectors in
calculation and not measured postselection angle in this range for
the saturation of CCD. In the previous work, the amplified factor
presents a valley near the Brewster angle on
reflection~\cite{Luo2011}. From the above analysis, we can select an
appropriate amplified factor in weak measurements under the
condition of fixed coupling strength, and so the high measurement
precision can be obtained. Note that the improvement of the
signal-to-noise ratio (SNR) is of great interest in precision
metrology. Recently, Jordan $et$ $al.$ have done great work to
optimize the SNR of a beam-deflection measurement with
interferometric weak values~\cite{Starling2009}. We think that how
to optimize the SNR for measuring photonic SHE with the
corresponding preselection and postselection states will be an
interesting work in the future.

In conclusion, we have used optimal preselection and postselection
in weak measurements to detect the photonic spin Hall effect (SHE)
for obtaining the high measurement precision. We have proved that
there exists the maximum weak value and amplified shift when the
overlap of preselection and postselection states are chosen as a
corresponding optimal value. We have also considered the orthogonal
condition of preselection and postselection states and revealed that
the weak value and amplified displacements of photonic SHE can not
be arbitrarily large. These findings provide new insight into weak
measurements and have possible applications in precision metrology.

We are sincerely grateful to the anonymous referee, whose comments
led to significant improvement of our paper. This research was
partially supported by the National Natural Science Foundation of
China (Grant No. 61025024 and No. 11274106) and Hunan Provincial
Innovation Foundation for Postgraduate (Grant No. CX2013B130).

\end{document}